\documentclass[journal]{IEEEtran}
\usepackage{cite}
\usepackage{amsmath}
\usepackage{makecell}
\usepackage{graphicx}
\usepackage{adjustbox}
\usepackage[dvipsnames]{xcolor}
\usepackage[T1]{fontenc}
\usepackage{multirow}

\ifCLASSINFOpdf
\else
\fi
\begin{document}
%
\title{Multi-Task Learning for Interpretable Weakly Labelled Sound Event Detection}
%
%
%
\author{Soham~Deshmukh,
        Bhiksha~Raj, Rita Singh
\thanks{S. Deshmukh is with Department of Electrical and Computer Engineering, Carnegie Mellon University}
\thanks{B. Raj is Professor at Language Technologies Institute, School of Computer Science, Carnegie Mellon University}
\thanks{R. Singh is Associate Research Professor at Language Technologies Institute, School of Computer Science, Carnegie Mellon University}
}
\maketitle

\begin{abstract}
Weakly Labelled learning has garnered lot of attention in recent years due to its potential to scale Sound Event Detection (SED) and is formulated as Multiple Instance Learning (MIL) problem. This paper proposes a Multi-Task Learning (MTL) framework for learning from Weakly Labelled Audio data which encompasses the traditional MIL setup. To show the utility of proposed framework, we use the input Time-Frequency representation (T-F) reconstruction as the auxiliary task. We show that the chosen auxiliary task de-noises internal T-F representation and improves SED performance under noisy recordings. Our second contribution is introducing two step Attention Pooling mechanism. By having 2-steps in attention mechanism, the network retains better T-F level information without compromising SED performance. The visualisation of first step and second step attention weights helps in localising the audio-event in T-F domain. For evaluating the proposed framework, we remix the DCASE 2019 task 1 acoustic scene data with DCASE 2018 Task 2 sounds event data under 0, 10 and 20 db SNR resulting in a multi-class Weakly labelled SED problem. The proposed total framework outperforms existing benchmark models over all SNRs, specifically 22.3 \%, 12.8 \%, 5.9 \% improvement over benchmark model on 0, 10 and 20 dB SNR respectively. We carry out ablation study to determine the contribution of each auxiliary task and 2-step Attention Pooling to the SED performance improvement. The code is publicly released.
\end{abstract}


%
\IEEEpeerreviewmaketitle

\section{Introduction}
%
%
%
%
\IEEEPARstart{T}{he} goal of Sound Event Detection (SED) is to determine the presence, nature and temporal location of  sound events in audio signals. This is usually accomplished using machine learning and signal processing algorithms, and is of great added benefit to applications like wearable devices \cite{wearable_devices}, mobile robots \cite{mobile_robots} and public security \cite{public_security}.

Many SED algorithms rely on strongly labelled data \cite{SLD_1}\cite{SLD_2}\cite{SLD_3} for training in order to perform accurate audio event detection. Here the term \textit{strongly labelled} refers to audio in which events are annotated with their corresponding onset and offset time. Usually, only the audio segments within the onset-offset time boundaries are used as training data, while segments outside the annotated onset and offset time boundaries are considered to be non-target events.  However, producing strongly labelled data for SED is quite expensive in terms of the expertise, time and human resources required for the purpose. For instance, the annotations produced are often restricted to minutes for every few hours \cite{SLD_3} of actual data and time spent listening to it. To address the scalability problem associated with generating the strongly labelled data, researchers have developed methods to train SED models from unstructured publicly available multimedia data. In this setting, the annotation generally takes the form of weak labels. The term 'weak' information or label refers to information which simply indicates presence or absence of particular events in the video or audio and does not provide any information with respect to number of audio events, duration of events and time localisation of events in the audio or video. Anurag et al. \cite{Anurag_WLD} proposed a way of training SED models based of weakly labelled data, where event detection is framed as Multiple-Instance Learning (MIL) problem \cite{MIL} and labels are available for a 'bag' of instances and not for individual instances. Specifically for SED, this results in every audio clip being considered as a 'bag' of instances, where the individual instances are the segments of the audio clip and clip level event information (label) is available. 

\begin{figure*}[!t]
    \centering
    \includegraphics[width=7in, height = 1.5in]{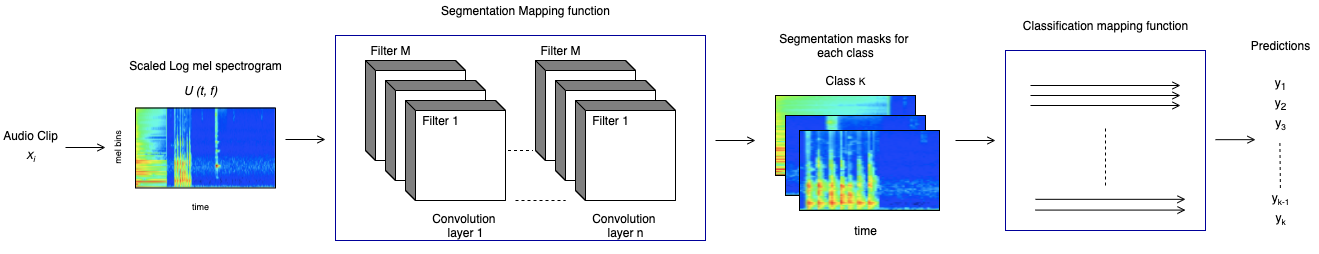}
    \centering
    \caption{Multiple-Instance Learning problem setup for Weakly Labelled SED. The setup is divided into three parts: input T-F representation, Segmentation Network and Classification Network, which outputs audio clip level sound event presence probabilities.}
    \label{fig:basic_problem_setup}
    \hfil
\end{figure*}

A specific type of MIL formulation which provides benchmark performance is neural MIL formulation. In neural MIL formulation, the model consists of two sequential components: a temporal information aggregator followed by a pooling mechanism. In short, the first component of the model (neural networks) produces temporal predictions which are then aggregated by the second half of the model, usually a pooling operator to produce audio clip level predictions. The temporal predictions are generally generated by encoding audio using a Convolution Neural Networks (CNN) \cite{Lecun} based architecture. The following pooling mechanism generally takes form of Global Max Pooling (GMP), Global Average Pooling (GAP), \cite{adaptive_pooling}, Global Weighted Rank Pooling (GWRP) \cite{gwrp}, Attention pooling \cite{atrous} to get audio level predictions. Depending on how the pooling is performed, has a significant performance effect on both the SED and each class's intermediate Time-Frequency (T-F) representation obtained. GMP and GAP consistently underestimate and overestimate the sound source time predictions respectively, and to overcome this problem smart and adaptive pooling mechanisms like Adaptive pooling \cite{adaptive_pooling}, Global Weighted Rank Pooling (GWRP) \cite{gwrp}, Attention pooling \cite{atrous} were developed. However, the developed pooling mechanisms still lacks the granularity of prediction required for inference of each audio event in T-F domain and is sometimes known to be unstable with Binary Cross Entropy loss usually used for multi-class WLSED \cite{atrous}.

To make the MIL setup more flexible and combat the above mentioned challenges, we propose a Multi-Task Learning (MTL) framework which uses a two-step attention pooling mechanism and signal reconstruction based auxiliary task. The chosen auxiliary task, results in providing a cleaned T-F representation of audio events, which can be further used for source separation. The two central contributions of the paper are:
\begin{itemize}
    \item Mutli-Task Learning formulation for WLSED: The paper introduces Multi-Task Learning setup for Weakly labelled Sound Event Detection (WLSED) where MIL based SED is the primary task and is assisted by an auxiliary task. The auxiliary task can be any secondary task whose labels are implicitly available or requires external meta-information. We choose input T-F signal reconstruction as auxiliary task as it does not require external labels or data. The input T-F reconstruction auxiliary task forces CNN encoded representation to retain and enhance each sound source’s temporal presence in the audio clip. Ablation studies are performed to quantify the effects of chosen auxiliary task to increased performance SED performance obtained. To the best of our knowledge, this is the first work formulating Multi-Task learning for Weakly Labelled SED data.
    \item Two-step Attention Pooling mechanism: The paper introduces a stable and interpretable two-step Attention Pooling mechanism along with MTL formulation. This helps in bringing intermediate T-F representation obtained from the initial CNN network to localised predictions of audio events in both temporal and frequency domain. The first step of attention operates over Mel bins and learns a contribution of each Mel bin to each time step for each audio event. The second step of attention operates over time and learns the contribution of audio events at each time step. The audio clip level predictions are obtained by summing up temporal contributions made by each audio event. This particular way of breaking the attention into two steps helps the network to easily focus on particular parts of inputs and localise in both time-frequency domain without making the training unstable and leads to benchmark performance.
\end{itemize}

The code is publicly realised and open-sourced. The paper is structured as follows: Section \ref{section:related_work} introduces Multiple-Instance Learning setup and previous work in Weakly labelled SED. Section \ref{section:proposedsolution} introduces Multi-Task Learning formulation with SED as the primary task. Section \ref{section:proposedsolution_classification} goes into the details of proposed 2-step Attention Mechanism. Section \ref{section:experiments} contains the experiment details about dataset, feature extraction and network details. Section \ref{section:results} contains the Weakly labelled SED results under different SNR, Ablation test results and End to End interpretable visualisation. Section \ref{section:futurework} contains future work and directions to build on this work, followed by Conclusion in Section \ref{section:conclusion}

\section{Problem setup and Related work \label{section:related_work}}
Weakly labelled learning in context of SED is generally formulated as Multiple Instance Learning (MIL) \cite{MIL} problem where a single binary class label is assigned to a collection (bag) of similar training examples (instances). The overall learning setup used for WLSED is show in Fig. \ref{fig:basic_problem_setup}
\subsection{MIL formulation for SED}
Let the raw audio be represented as $\{{x_i}\}_{i = 1}^T$, where $x_i$ is an individual frame out of the T frames making up the audio. Let the features extracted from raw audio $\{{x_i}\}_{i = 1}^T$ be represented by $\{{\hat{x_i}}\}_{i = 1}^T$. The extracted features constitute a bag B = $\{\hat{x_i}\}_{i = 1}^T$. MIL assumption states that the weak labels of bag B are $y = \text{max}_i\{y_i\}_{i = 1}^T$, where $y_i$ is the strong label corresponding to feature $\hat{x_i}$. Thus training pairs of Weakly labelled data consists of $\{ B, y\}$. The mapping of frame level features to audio level event prediction is learned by neural networks in Neural MIL. The neural network can be divided into two smaller sequential neural networks. The first network learns a function $g_1$ which generates a T-F segmentation mask. Ideally, the output generated by this network is supposed to be T-F segmentation masks for each audio event. The paper will refer to the first network as \lq Segmentation network\rq. This is followed by a second network which learns mapping $g_2$ between T-F segmentation masks to probabilities of corresponding sound events. The output of the second network is audio clip level class probabilities. As the second network's role is to perform classification of T-F representation into appropriate audio events, the paper will refer to this second network as \lq Classification Network\rq.

\subsection{Segmentation Network \label{section:Relatedwork_segmentation}}
The initial work \cite{Anurag_WLD}, proposes SVM max margin formulation and CNN based architecture as two alternatives for learning the segmentation mapping. To improve the quality of segmentation maps learned, \cite{Florian}\cite{DeepCNN_anurag} used a Time-Distributed CNN (TD-CNN) followed by a Global Max Pooling (GMP) to pick out location of relevant temporal events. In most of the relevant work, log Mel spectrograms are used as features to represent raw audios. \cite{environmental_sound_classification, esc_deepcnn, largescale_audioclassification} shows using log mel spectrogram representation of input followed by CNN based architecture leads to improved performance compared to machine learning methods like SVM. To better capture temporal event information present in audio, \cite{gcnn} used convolutional Recurrent Neural Network (CRNN). CRNN consists of CNN layers which operate on log Mel spectrogram input to extract relevant source level features followed by Bidirectional Recurrent Neural Network (Bi-RNN) to capture the temporal context information between frames in T-F representation. Recently \cite{atrous} proved that size of receptive field is more important than performing intermediate local pooling, and used atrous CNNs \cite{atrous_CNN_original} with global attention pooling layer to achieve benchmark performance. In this paper, we intend to achieve better explainability in the T-F domain without compromising SED performance, by reconstruction based auxiliary task and 2-Step Attention Pooling. The proposed methodology helps in denoising the internal T-F representations and provides audio event localisation in temporal and frequency domain, along with achieving benchmark performance in WLSED.

\subsection{Classification Network \label{section:Relatedwork_classification}}
Classification mapping takes intermediate T-F representation as input and employes pooling operators such as Global Max Pooling (GMP), Global Average Pooling (GAP) \cite{gap}, Global Weighted Rank Pooling (GWRP) \cite{gwrp}, global attention pooling \cite{attentionmodel}\cite{atrous} or other poolings \cite{framecnn}\cite{gaussianfilter}\cite{adaptive_pooling} and even fully connected layers to predict the presence specific of sound events in the audio. The GMP operator is known to under predict the sound events as the operator only takes into account the most prominent feature while ignoring others. On the other hand, GAP is consistently known to over predicts the sounds events \cite{gwrp}. To address this challenge, \cite{gwrp} uses GWRP operator which can be seen as an extension of both GAP and GMP. The GWRP operator converts to GMP when r = 0 and GAP when r = 1, where r is a hyperparameter that varies according to frequency of occurrence of sound events. However, any type of global pooling can only predict the time domain segmentation, but not the T-F segmentation. To make the pooling operator more flexible and aware of each contributing location in the T-F domain, an attention mechanism was proposed \cite{attentionmodel}\cite{atrous}. Attention mechanism is more flexible and weighs each contributing location in T-F output to form the final predictions. However, training using attention proposed in \cite{attentionmodel}\cite{atrous} over the entire K $\times$ T $\times$ F (where K is sound events, T is time frames and F is frequency bins) is unstable with commonly used Binary Cross Entropy loss. Both the challenges of fine-grained sound event contribution in the T-F domain and instability are addressed by two-step Attention Pooling (2AP) mechanism proposed in this paper.

\subsection{Multi-Task Learning}
Multi-Task Learning has been associated with many other names, some of the common ones are joint learning, learning to learn, and learning with auxiliary tasks. Multi-Task Learning (MTL) is a type of inductive transfer, where this inductive bias generally takes the form of an auxiliary task, which forces the network to learn representation to jointly solve more than one task. This generally leads to solutions that generalise better \cite{mtl} and is shown to work across many application of machine learning natural language processing, speech recognition and computer vision \cite{mtl_speech}\cite{mtl_vision}\cite{mtl_nlp}. 

In audio domain, MLT has been recently applied to jointly learn features for multiple speech classification tasks: speaker identification, emotion classification, and automatic speech recognition \cite{pase}\cite{ravanelli} in which the shared network learns representation to solve all the downstream tasks equally well. Our MTL setup employs a different formulation, where rather than multiple tasks, we employ an auxiliary task which doesn't require additional data or labels and the auxiliary task's performance is not of interest. The key is selecting appropriate auxiliary tasks for Multi-Task Learning, if incorrectly selected the auxiliary task can hurt the performance of the primary task. To the best of our knowledge, this is the first work which formulates Multi-Task Learning for Weakly Labelled SED data and determines appropriate auxiliary tasks for the same.

\begin{figure*}
    \centering
    \includegraphics[width=7 in, height = 3.5in]{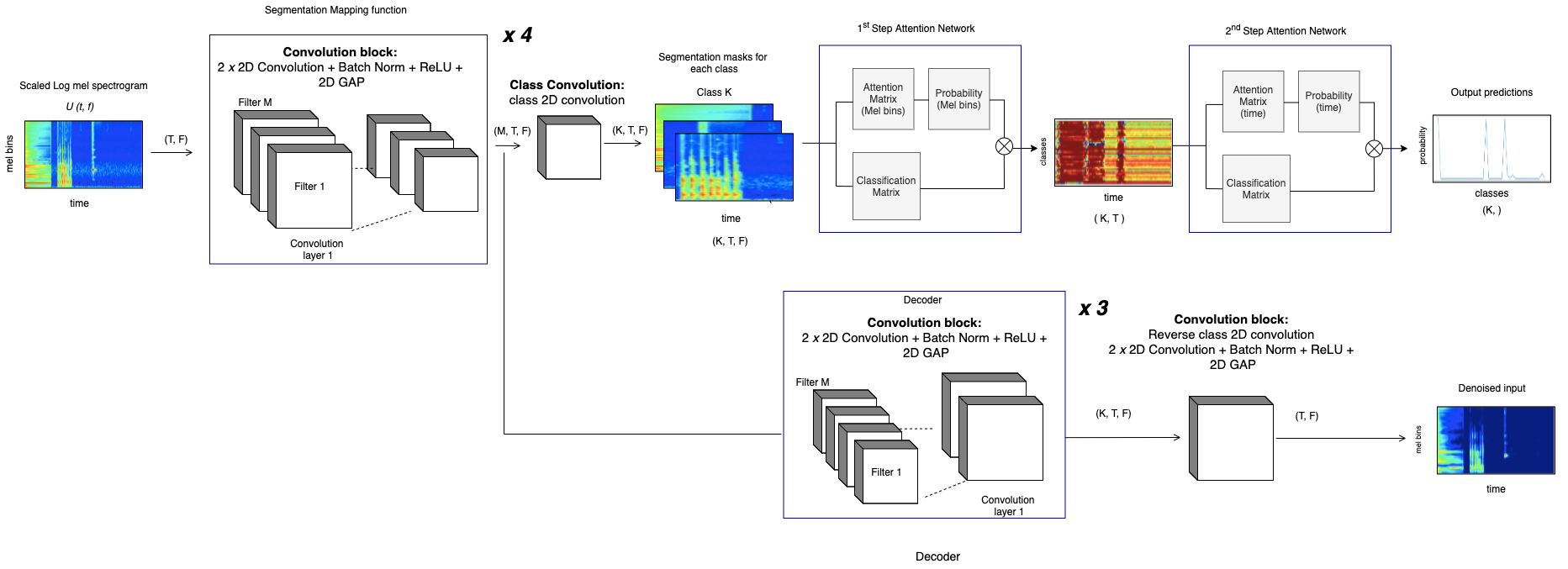}
    \caption{Multi-Task Learning with T-F reconstruction as auxiliary task. The upper part of network is the primary WLSED task, with main block as: input T-F representation, Segmentation Network and Classification Network. The lower part of network is of the auxiliary task which is reconstruction of internal T-F representation.}
    \label{fig:MTL_setup}
    \hfil
\end{figure*}

\section{Proposed Methodology \label{section:proposedsolution}}
This section contains the details of Multi-Task Learning formulation for SED, its corresponding segmentation mapping network $g_1$, classification mapping network $g_2$, and the auxiliary T-F reconstruction auto-encoder task. This paper's Multi-Task Learning formulation is depicted in Fig \ref{fig:MTL_setup}

\subsection{Multi-Task Learning SED formulation \label{section:proposedsolution_formulation}}
In this section we formalise the Multi-Task Learning setup for Weakly Labelled SED. Let the raw audio be represented as $X = \{{x_i}\}_{i = 1}^T$ where $x_i \in R$ and $i$ indicates the specific frame in of total frames T. Features are extracted from raw audio and brought to a T-F form: $\hat{X} = \{{\hat{x_i}}\}_{i = 1}^T$ where $\hat{x_i} \in R^F$ where F is the number of frequency bins used to represent a frame of audio. Every bag of frames B = $\{\hat{x_i}\}_{i = 1}^T$ has a weak label $y \in R^K$ where K is the number of audio events. Then the weakly labelled training data in terms of bags is represented as:
\begin{equation}
    B_j = (\{\hat{x_i}\}_{i = 1}^T, y)\vert_{j = 0}^N
\end{equation}
where N are the number of examples or data points. 

As the primary task in our setup of Multi-task learning framework is Sound Event Detection, the first part of network for SED task focuses on learning a segmentation mapping $g_1(.)$ where:
\begin{equation}
    g_1 : \hat{X} \mapsto Z
\end{equation}
The segmentation network maps the feature set $\{{\hat{x_i}}\}_{i = 1}^T$ to $Z = \{z_i\}_{i = 1}^T$ where $z_i \in R^{K \times F}$ and $K$ is the number of audio events. The second part of SED task is network which classifies $\{z_i\}_{i = 1}^T$ to $P = \{ p_i \}_{i = 1}^K$ where $P \in R^K$. The network learns a mapping: 
\begin{equation}
    g_2 : Z \mapsto P
\end{equation}
where $g_2$ maps each classes T-F segmentation to their presence probabilities of $k^{th}$ event known as $p_k$

The auxiliary task in MTL setup is reconstruction of input T-F representation. For reconstruction, an autoencoder structure will be used. The first part of autoencoder is an encoder network which compacts data into an intermediate T-F representation. The aim of compacted features is to allow reconstruction of data with minimal error. The encoder learns a function $g_3(.)$ where $g_3 : \hat{X} \mapsto Z$. 

We make an assumption that audio's T-F representation $\{{\hat{x_i}}\}_{i = 1}^T$ can be completely explained as a linear or non-linear combination of each audio event's independent T-F representation. This assumption allows the network to have a shared encoder for both SED and auxiliary reconstruction tasks. From now on we will represent $g_1(.) = g_3(.) = g(.)$ as the shared segmentation mapping function, such that $g : \hat{X} \mapsto Z$. The shared encoder performs compaction in the number of filters and keeps the T-F dimensions untouched.

The second part of the auxiliary task is a decoder network which learns a mapping $g_4$ such that $g_4: Z \mapsto \bar{X}$ where $\bar{X}$ is the reconstructed T-F representation. Specifically:
\begin{equation}
    \{\bar{x_i}\}_{i = 1}^T = g_4(\{z_i\}_{i = 1}^T)
\end{equation}
Here the mapping function $g_4(.)$ is ideally such that: 
\begin{equation}
    g_4(g(.)) = g(g_4(.)) = I
\end{equation}
To force the above relation in auxiliary task, we introduce loss function $\mathcal{L}_2$ to minimise the difference between true T-F representation $\{\hat{x_i}\}_{i = 1}^T$ and predicted T-F representation $\{\bar{x_i}\}_{i = 1}^T$ of audio.

To solve for the SED task, the network should learn $g(.)$ shared mapping such that the mask $\{z_i\}_{i = 1}^T$ should accurately segment each audio event and classification mapping $g_3(.)$ should map it to correct audio event. Let the loss to enforce these conditions for the primary task be $\mathcal{L}_1$. The functions $g(.), g_2(.), g_4(.)$ will be expressed using Neural Networks. In terms of Neural Network terminology, this is equivalent to learning a weights W = $[w, w_2, w_4]$ where $w, w_2, w_4$ are weights corresponding to each function $g(.), g_2(.), g_4(.)$ respectively. The optimisation problem can be framed in terms of these weights W over all data points as:
\begin{equation}
    \underset{W}{\text{min}}\; \mathcal{L}_1(P, y \vert w, w_4) + \alpha \mathcal{L}_2(\{\bar{x_i}\}_{i = 1}^T, \{\hat{x_i}\}_{i = 1}^T \vert w, w_2)
\end{equation}
The two-component loss function will be to referred as $L(W)$ and parameter $\alpha$ accounts for scale difference between losses $\mathcal{L}_1$ and $\mathcal{L}_2$. It also helps in adjusting weightage given to auxiliary task relative to primary task to guide learning of weights.

\subsection{Shared Segmentation Network \label{section:proposedsolution_segmentation}}
The Segmentation mapping here processes the audio clip and converts it into a processed T-F (Time Frequency) representation for each class. The audio clip $X_i$ is first converted into a T-F representation $\hat{X} = \{{\hat{x_i}}\}_{i = 1}^T$. Here $\hat{X} = \{{\hat{x_i}}\}_{i = 1}^T$ is log mel spectrogram. CNN based network is used here for modelling mapping $g(.)$. The CNN based network has multiple CNN blocks. Each CNN block consists of three subparts. In first part, features $\hat{X} = \{{\hat{x_i}}\}_{i = 1}^T$ is processed using 2 repeating units of 2-dimensional linear convolution, Batch Normalisation \cite{batch_norm} and ReLU \cite{relu} nonlinearity. This is followed by a 2-dimensional Average pooling operator with appropriate stride and padding to main original T-F dimensions of audio. Batch Normalisation helps to stabilise training by reducing the layer’s internal covariate shift and recent work suggests it makes the optimisation landscape significantly smoother which results in the stable behaviour of gradients \cite{batch_norm_justification}. This 2 unit CNN block is repeated to form the segmentation mapping section of the network. 

The segmentation network also acts as the encoder of the auto-encoder framework and has to jointly encode features relevant for audio event detection and audio T-F representation reconstruction. Having a common encoder forces the network to learn a shared representation, helps the network to exploit commonalities and differences across SED and T-F reconstruction and enables the network to generalise better on our original task. The Multi-Task Learning (MLT) setup using an auxiliary task to introduce an inductive bias, which will cause the network to prefer solutions that generalise better \cite{mtl_overview}.  This will result in improved learning and predictive power for SED. The MLT setup used here is a hard parameter sharing instead of soft parameter sharing which greatly reduces the risk of overfitting \cite{mtl_informationtheory}. 

\subsection{Classification Network \label{section:proposedsolution_classification}}
The modelling choice for classification mapping results in different intermediate T-F representation of audio. The traditional choices for modelling classification mapping are Global Max Pooling \cite{gap}, Global Average Pooling \cite{gmp} and Global Weighted Rank Pooling \cite{gwrp_original} which results in compromise between time level precision and classification performance. We instead propose a 2-step Attention Pooling mechanism which retains interpretability in the T-F domain and provides better SED classification performance. The two-step Attention Pooling mechanism converts segmentation mapping $\{z_i\}_{i = 1}^T$ into audio level predictions $P$.

The 2-step Attention Pooling takes input $\{z_i\}_{i = 1}^T$ of size $K \times F \times T$ where K, F, T denotes the number of classes, frequency bins and time frames respectively, where the first step operates on $Z = \{z_i\}_{i = 1}^T$ and produces $Z_{p1}$ as the intermediate output, followed by second step which operates on $Z_{p1}$ and produces $Z_{p2}$ as the final output indicating presence probability of different sound events in the audio clip. In the first step, two independent linear neural networks operate on $Z$ to produce an attention weight $Z_{a1}$ and an intermediate classification output $Z_{p1}$ respectively. Softmax is used to ensure attention weights sum up to 1 along the frequency dimension, to produce normalised attention weights $\widehat{Z}_{a1}$. The normalised attention weights are used to weigh the intermediate classification outputs to produce $Z_{p1}$ of size K $\times$ T with squashed frequency dimension.
\begin{equation}
    Z_{a1} = \sigma(ZW_{a1}^T + b_{a1})
\end{equation}
\begin{equation}
    Z_{c1} = ZW_{c1}^T + b_{c1}
\end{equation}
\begin{equation}
    \widehat{Z}_{a1} = \frac{e^{Z_{a1}}}{\sum_{i = 1}^F e^{Z_{a1}}}
\end{equation}
$\hat{Z}_{a1}$ are the probabilities to weigh the classification branch results:
\begin{equation}
    Z_{p1} = \sum_{i = 0}^F Z_{c1} \cdot \hat{Z}_{a1}
\end{equation}
Subsequently, $Z_{p1}$ is passed as input to the second step attention pooling. The second step attention pooling performs the same general process as the first step attention with the difference of operating Softmax and squashing on the time dimension. 
\begin{equation}
    Z_{a2} = \sigma(Z_{p1}W_{a1}^T + b_{a1})
\end{equation}
\begin{equation}
    Z_{c2} = Z_{p1}W_{c2}^T + b_{c2}
\end{equation}
\begin{equation}
    \widehat{Z}_{a2} = \frac{e^{Z_{a2}}}{\sum_{t = 1}^T e^{Z_{a2}}}
\end{equation}
\begin{equation}
    Z_{p2} = \sum_{t = 0}^T Z_{c2} \cdot \hat{Z}_{a2}
\end{equation}
$Z_{p2} \in (0,1)$ is of size K and denotes the probability of each sound event k $\in$ K in the audio clip. By breaking the attention into two steps, it makes the pooling more interpretable by answering the questions of what frequency bins and what time steps contributes to which audio events by visualising normalised attention weights $\widehat{Z}_{a1}, \widehat{Z}_{a2}$ and each step's attention output $Z_{p1}, Z_{p2}$. By adding $\sigma$ in both step's intermediate classification output, it ensures the output is bounded between 0 and 1, leading to stable training with Binary Cross Entropy loss used for computing error between the predicted probability of audio events $P$ and true weak labels $y$. The quantitative performance comparison between traditional attention pooling and 2-step attention is done in the results section. 

\subsection{Decoder Network \label{section:proposedsolution_decoder}}
The decoder of the auxiliary-task takes $Z = \{z_i\}_{i = 1}^T$ as input and tries to reconstructs $\{\hat{x_i}\}_{i = 1}^T$. The decoder uses CNN based network for down-sampling number of filters from K (Number of classes). The CNN based network has multiple CNN blocks, where each subblock consists of three subparts. In first part, features $\hat{X} = \{{\hat{x_i}}\}_{i = 1}^T$ is processed using 2 repeating units of 2-dimensional linear convolution, Batch Normalisation \cite{batch_norm} and ReLU \cite{relu} nonlinearity. This is followed by a 2-dimensional Average pooling layer with appropriate stride and padding to main original T-F dimensions of audio. The general architecture of decoder closely follows the shared encoder architecture, with exception that the number of filters are decreased from one block to another in the decoder, while they are increased in the encoder architecture. For training the weights, the loss used for the auxiliary task is Mean Squared Error (MSE) which computes error between T-F representation $\{\hat{x_i}\}_{i = 1}^T$ and reconstructed T-F representation $\{\bar{x_i}\}_{i = 1}^T$

\begin{table}[!t] 
\renewcommand{\arraystretch}{1.3} 
\centering
\caption{Shared Segmentation Network} 
\label{table:segmentation_network}
\begin{tabular}{c |c}
\hline
Layers & Output size \\ 
\hline
log mel spectrogram & 1 $\times$ 311 $\times$ 64 \\
Conv2D: \{K = 3, C = 32\}, BN, ReLU $\times$ 2 & 32 $\times$ 311 $\times$ 64 \\
2D Average pooling \{K = 3, S = 1, P = 1\} & 32 $\times$ 311 $\times$ 64 \\
Conv2D: \{K = 3, C = 64\}, BN, ReLU $\times$ 2 & 64 $\times$ 311 $\times$ 64 \\
2D Average pooling \{K = 3, S = 1, P = 1\} & 64 $\times$ 311 $\times$ 64 \\
Conv2D: \{K = 3, C = 128\}, BN, ReLU $\times$ 2 & 128 $\times$ 311 $\times$ 64 \\
2D Average pooling \{K = 3, S = 1, P = 1\} & 128 $\times$ 311 $\times$ 64 \\
Conv2D: \{K = 3, C = 256\}, BN, ReLU $\times$ 2 & 256 $\times$ 311 $\times$ 64 \\
2D Average pooling \{K = 3, S = 1, P = 1\} & 256 $\times$ 311 $\times$ 64 \\
Conv2D: \{K = 1, C = 41\} & 41 $\times$ 311 $\times$ 64 \\
\hline
\end{tabular}
\end{table}

\begin{table}[!t] \renewcommand{\arraystretch}{1.3} 
\centering
\caption{Classification Network} 
\label{table:classificaion_network}
\begin{tabular}{c |c}
\hline
Layers & Output size \\ 
\hline
class T-F representation & 41 $\times$ 311 $\times$ 64 \\
$1^{st}$ step Attention Pooling & 41 $\times$ 311 \\
$2^{nd}$ step Attention Pooling & 41 \\
\hline
\end{tabular}
\end{table}

\begin{table}[!t] \renewcommand{\arraystretch}{1.3} 
\centering
\caption{Decoder Network} 
\label{table:decoder_network}
\begin{tabular}{c |c}
\hline
Layers & Output size \\ 
\hline
Internal T-F representation & 256 $\times$ 311 $\times$ 64 \\
Conv2D: \{K = 3, C = 128\}, BN, ReLU $\times$ 2 & 128 $\times$ 311 $\times$ 64 \\
2D Average pooling \{K = 3, S = 1, P = 1\} & 128 $\times$ 311 $\times$ 64 \\
Conv2D: \{K = 3, C = 64\}, BN, ReLU $\times$ 2 & 64 $\times$ 311 $\times$ 64 \\
2D Average pooling \{K = 3, S = 1, P = 1\} & 64 $\times$ 311 $\times$ 64 \\
Conv2D: \{K = 3, C = 32\}, BN, ReLU $\times$ 2 & 32 $\times$ 311 $\times$ 64 \\
2D Average pooling \{K = 3, S = 1, P = 1\} & 32 $\times$ 311 $\times$ 64 \\
Conv2D: \{K = 3, C = 1\}, BN, ReLU $\times$ 2 & 311 $\times$ 64 \\
\hline
\end{tabular}
\end{table}

\section{Experiments \label{section:experiments}}
\subsection{Dataset \label{section:experiments_dataset}}
The dataset is made by mixing DCASE 2019 task 1 of Acoustic scene classification \cite{dcase1} and DCASE 2018 task 2 of General purpose Audio tagging \cite{dcase2}. The dataset creation methodology used is the same as \cite{gwrp}. However, we use new DCASE 2019 Task 1 instead of DCASE 2018 used in \cite{gwrp}. We use the DCASE 2019 variant as it extends the TUT Urban Acoustic Scenes 2018 with other 6 cities to a total of 12 large European cities, making a total dataset more diverse and challenging. Specifically, TUT 2018 Urban Acoustic Scenes dataset contains recordings from Barcelona, Helsinki, London, Paris, Stockholm and Vienna, to which TAU 2019 Urban Acoustic Scenes dataset adds Lisbon, Amsterdam, Lyon, Madrid, Milan, and Prague. This provides for the background or environmental noise needed to simulate real world audio event background. The DCASE 2018 task 2 provides annotated audio clips associated with one of the 41 events like \lq Flute\rq, \lq Gunshot\rq and \lq Bus\rq from Google Audioset Ontology \cite{Audioset}. On the other hand, DCASE 2019 task 1 contains 10 sec clips from 10 different scenes like “airport”, “metro station” and “urban park” to name a few. From DCASE 2018 task 2, two second clips are extracted and mixed with 10 second background noise from DCASE 2019 task 1 for SNR 0 dB, 10 dB and 20 dB. Each audio clip contains three non-overlapping audio events. For each SNR, the 8000 mixed audio clips are divided into 4 cross-validation folds. The mixed audio clips are single channel with a sampling rate of 32 kHz and the mixing procedure can be found in \cite{gwrp}.

\begin{table*}[!t] \renewcommand{\arraystretch}{1.3} 
    \centering
    \caption{Weakly Labelled SED Results} 
    \label{table:sed_results}
    \begin{tabular}{ c | l  l  l | l  l  l | l  l  l }
        \hline
        \multirow{2}{*}{Networks} 
            & \multicolumn{3}{c}{SNR 20 dB} 
                & \multicolumn{3}{c}{SNR 10 dB}
                    & \multicolumn{3}{c}{SNR 0 dB} \\ 
        & micro-P & macro-P & AUC & micro-P & macro-P & AUC & micro-P & macro-P & AUC \\ \hline
        GAP &  0.5067& 0.6127& 0.9338 & 0.4291 & 0.5390 & 0.5390 & 0.3295 & 0.4093 & 0.8694 \\
        GMP &  0.5390&  0.5186&  0.8497&  0.5263&  0.5023&  0.8422&  0.4640&  0.4441&  0.8189\\
        GWRP \cite{gwrp} &  0.7018& 0.7522&  0.9362&  0.6538&  0.7129&  0.9265&  0.5285&  0.6084&  0.8985\\
        Atrous AP \cite{atrous} &  0.7391& 0.7586&  0.9279&  0.6740&  0.7404&  0.9211&  0.5714&  0.6341&  0.9014\\
        2APAE &  \bf{0.7829}&  \bf{0.7645}&  \bf{0.9390}&  \bf{0.7603}&  \bf{0.7486}&  \bf{0.9343}&  \bf{0.6986}&  \bf{0.6892}&  \bf{0.9177}\\
    \hline
    \end{tabular}
\end{table*}

\begin{table*}[!t] \renewcommand{\arraystretch}{1.3} 
    \centering
    \caption{Ablation Study} 
    \label{table:ablation_study_results}
    \begin{tabular}{ c | l  l  l | l  l  l | l  l  l }
        \hline
        \multirow{2}{*}{2APAE} 
            & \multicolumn{3}{c}{SNR 20 dB} 
                & \multicolumn{3}{c}{SNR 10 dB}
                    & \multicolumn{3}{c}{SNR 0 dB} \\ 
        & micro-P & macro-P & AUC & micro-P & macro-P & AUC & micro-P & macro-P & AUC \\ \hline
        $\alpha$ = 0.0 &  0.7772& \bf{0.7648}& 0.9389 & 0.7430 & 0.7383 & 0.9299 & 0.6937 & 0.6807 & 0.9144 \\
        $\alpha$ = 0.001 &  \bf{0.7829}&  0.7645&  \bf{0.9390}&  \bf{0.7603}&  \bf{0.7486}&  \bf{0.9343}&  \bf{0.6986}&  \bf{0.6892}&  \bf{0.9177}\\
        $\alpha$ = 0.01 &  0.7637& 0.7464&  0.9333&  0.7428&  0.7278&  0.9277&  0.6792&  0.6689&  0.9114\\
    \hline
    \end{tabular}
\end{table*}

\subsection{Evaluation Metric \label{section:experiments_evaluation}}
The metric used to evaluate the proposed framework and benchmark models is Area Under the Receiver Operating Characteristic Curve (ROC AUC) \cite{roc_auc}, micro Precision and macro Precision. The metrics are chosen such that they are threshold independent and characterise the performance of the network in both balanced and imbalanced class settings. The metrics are defined as follows:
\begin{enumerate}
    \item ROC AUC: A receiver operating characteristic curve plots true positive rate (TPR) vs false positive rate (FPR). The area under the ROC curve is computed which summaries the ROC AUC curve. Using the AUC does not require manual selection of a threshold. Bigger AUC indicates better performance. A random guess has an AUC of 0.5
    \item Precision: The precision is intuitively the ability of the classifier not to label as positive a sample that is negative. 
    \begin{align}
        P = \frac{TP}{TP + FP}
    \end{align}
    There are different ways of adapting this metric to multi-label/multi-class classification. We use both \lq Micro\rq  and \lq Macro\rq  precision for evaluation \cite{52}. Micro Precision calculates the metrics globally by counting the total true positives, false negatives and false positives. On the other hand Macro Precision calculates metrics for each label, and finds their unweighted mean. Macro Precision does not take label imbalance into account and hence we consider Micro Precision to be a metric of primary importance here.
\end{enumerate}

\subsection{Feature Extraction \label{section:experiments_featureextraction}}
The raw data is converted to T-F representation by applying FFT with a window size of 2048 and overlap of 1024 between consecutive windows. This is followed by applying mel filter banks with 64 bands and converting it to log scale to obtain log mel spectrogram. This is used as input to the shared segmentation mapping network and is found to work well with neural networks \cite{gwrp, 48}. 

\subsection{Network \label{section:experiments_network}}
This subsection provides shared segmentation, classification and decoder network's detailed description along the training details. The segmentation network details are shown in Table \ref{table:segmentation_network}. It takes log mel spectrogram as input and employs a CNN block structure similar to VGG \cite{vdcnn}. The smallest unit of the network consists of a CNN layer of kernel size 3, Batch Normalisation and ReLU sequentially called CNN sub-block. This sub-block is repeated twice and is followed by 2D Average Pooling of stride 1 and padding 1 to form a CNN block. Empirically, 2D Max pooling and 2D Average Pooling showed similar performance and therefore 2D Average Pooling was arbitrarily chosen. The CNN block is repeated 4 times with increasing channel size (32, 64, 128, 256). This input is passed to the class convolution block as shown in Fig. \ref{fig:MTL_setup} to reduce it down to number of classes. Note that the Decoder network takes the input of class convolution as it's input and not the output of class convolution. This is determined empircally, as the performance of both is comparable and considering the input of class convolution saves an additional convolution block which would be necessary to upscale the class T-F representation in the decoder network.

The classification network (Table \ref{table:classificaion_network}), consists of 2 step attention pooling to summarize and reduce the class T-F representation to get audio clip level predictions. The 2-step attention pooling is described in Section \ref{section:proposedsolution_classification}. The Decoder network (Table \ref{table:decoder_network}) emulates the encoder network with decrease in the number of channels to reduce it to log mel spectrogram of input. Using the terminology of sub-block and block, the decoder consists of 3 CNN-blocks with decreasing channel size (256, 128, 64, 32). This is followed by a reversed class convolution block consisting of one CNN block with only one channel to reduce the T-F representation to log mel spectrogram. The entire network is trained end-to-end with a batch size of 24 and learning of 1e-3 using Adam optimiser \cite{adam} on 4 GPU cards with 12 GB RAM each. 

\section{Results \label{section:results}}
\subsection{Sound Event Detection Results \label{section:results_sed}} 
The proposed architecture is compared with traditional methods and current benchmark architecture for weakly labelled sound event detection specifically GMP, GAP, GWRP \cite{gwrp}, Attention pooling with Atrous convolution \cite{atrous}. The performance of different architectures are compared in Table \ref{table:sed_results} where the each metric is an average cross-validation score obtained across 4 folds. The important metric here is Micro Precision (micro-P), as it calculates metrics globally by counting the total true positives, false negatives and false positives. This is a better indicator of network performance as it takes into account class imbalance rather than simple unweighted averaging that macro Precision does. 2APAE (2 step Attention Pooling with Auto Encoder auxiliary task) has micro-Precision score of 0.7829, 0.7603 and 0.6986 on SNR 20, 10 and 0 dB respectively. The results show that the 2APAE network achieves the best score across all SNR's across all metrics. In terms of micro-Precision, 2APAE outperforms existing benchmark of Atrous AP (Atrous Attention Pooling \cite{atrous}) on SNR 20, 10 and 0 dB by 5.9\%, 12.8\% and 22.3\% respectively. 

The second benefit of breaking the attention into two steps apart from improving performance, is to provide stable training. The attention pooling used with atrous convolution resulted in overflow issues during training, where the final predictions probabilities when over 1 for some audio events or classes. This doesn't fit well with Binary Cross entropy as loss function which expects the input probabilities between 0 to 1. Empirically, adding a squashing function after attention hampers learning. By breaking the attention into two steps, it allows for the intermediate use of sigmoid which helps in ensuring the outputs don't explode above 1. 

\begin{table*}[!t]
\tiny \centering
\caption{Weakly Labelled SED audio event specific results for snr = 0} 
\label{table:class_sed_results_0}
\begin{tabular}{ c | c | c | c | c | c | c | c | c | c | c | c | c | c | c | c | c | c | c | c | c}
\hline
Model & \makecell{Guit\\ar} & \makecell{Appl\\ause} & \makecell{Ba\\rk} & \makecell{Bass\\drum} & \makecell{Bur\\ping} & \makecell{Bus} & \makecell{Cel\\lo} & \makecell{Chi\\me} & \makecell{Clar\\inet} & \makecell{Comp.\\keyb.} & \makecell{Cou\\gh} & \makecell{Cow\\bell} & \makecell{Double\\bass} & \makecell{Dra\\wer} & \makecell{Elec.\\piano} & \makecell{Fa\\rt} & \makecell{Finger\\snapp.} & \makecell{Fire\\work} & \makecell{Flu\\te} & \makecell{Glock\\ensp.}\\
\hline 
GAP & 0.549 & 0.848 & 0.477 & 0.161 & 0.508 & 0.168 & 0.361 & 0.626 & 0.289 & 0.502 & 0.384 & 0.447 & 0.199 & 0.212 & 0.251 & 0.386 & 0.409 & 0.36 & 0.286 & 0.539 \\
GMP & 0.517 & 0.539 & 0.53 & 0.535 & 0.426 & 0.145 & 0.378 & 0.406 & 0.466 & 0.356 & 0.208 & 0.872 & 0.275 & 0.077 & 0.31 & 0.393 & 0.623 & 0.322 & 0.384 & 0.889\\
GWRP & 0.728 & 0.933 & 0.742 & 0.242 & 0.741 & 0.254 & 0.511 & 0.766 & 0.449 & 0.587 & 0.629 & 0.768 & 0.262 & 0.296 & 0.349 & 0.652 & 0.514 & 0.517 & 0.418 & 0.893\\
AtrousAP & 0.72 & 0.956 & 0.782 & 0.169 & 0.804 & 0.2 & 0.562 & 0.767 & 0.502 & 0.685 & 0.756 & 0.781 & 0.17 & 0.214 & 0.187 & 0.691 & 0.734 & 0.566 & 0.318 & 0.902\\
2APAE & 0.869 & 0.942 & 0.865 & 0.82 & 0.849 & 0.572 & 0.71 & 0.633 & 0.542 & 0.59 & 0.628 & 0.921 & 0.579 & 0.386 & 0.552 & 0.569 & 0.907 & 0.579 & 0.473 & 0.907\\
2APAE e-3 & 0.792 & 0.951 & 0.839 & 0.812 & 0.874 & 0.627 & 0.669 & 0.606 & 0.503 & 0.699 & 0.631 & 0.94 & 0.59 & 0.403 & 0.453 & 0.562 & 0.941 & 0.565 & 0.535 & 0.807\\
2APAE e-2 & 0.759 & 0.943 & 0.787 & 0.789 & 0.81 & 0.605 & 0.677 & 0.637 & 0.485 & 0.68 & 0.632 & 0.916 & 0.563 & 0.377 & 0.522 & 0.589 & 0.867 & 0.61 & 0.522 & 0.853\\ \hline
\end{tabular}
\end{table*}
\begin{table*}[!t]
\tiny \centering
\begin{tabular}{ c | c | c | c | c | c | c | c | c | c | c | c | c | c | c | c | c | c | c | c | c }
\hline
\makecell{Gong} & \makecell{Gun\\shot} & \makecell{Harm\\onica} & \makecell{Hi-\\hat} & \makecell{Keys} & \makecell{Kno\\ck} & \makecell{Laugh\\ter} & \makecell{Me\\ow} & \makecell{Micro.\\oven} & \makecell{Oboe} & \makecell{Saxo\\phone} & \makecell{Scis\\sors} & \makecell{Shat\\ter} & \makecell{Snare\\drum} & \makecell{Squ\\eak} & \makecell{Tamb\\ourine} & \makecell{Tear\\ing} & \makecell{Tele\\phone} & \makecell{Trum\\pet} & \makecell{Violin\\fiddle} & \makecell{Writ\\ing} \\
\hline
0.34 & 0.473 & 0.698 & 0.717 & 0.384 & 0.42 & 0.396 & 0.3 & 0.193 & 0.288 & 0.477 & 0.456 & 0.527 & 0.344 & 0.174 & 0.512 & 0.357 & 0.272 & 0.514 & 0.474 & 0.377 \\
0.416 & 0.43 & 0.375 & 0.887 & 0.493 & 0.52 & 0.406 & 0.314 & 0.215 & 0.485 & 0.566 & 0.344 & 0.416 & 0.462 & 0.077 & 0.911 & 0.39 & 0.345 & 0.569 & 0.674 & 0.192 \\
0.576 & 0.645 & 0.851 & 0.847 & 0.624 & 0.543 & 0.585 & 0.548 & 0.367 & 0.495 & 0.654 & 0.545 & 0.684 & 0.513 & 0.207 & 0.866 & 0.552 & 0.49 & 0.664 & 0.594 & 0.52 \\
0.692 & 0.684 & 0.861 & 0.919 & 0.78 & 0.694 & 0.628 & 0.583 & 0.157 & 0.565 & 0.684 & 0.713 & 0.777 & 0.446 & 0.223 & 0.952 & 0.573 & 0.52 & 0.793 & 0.743 & 0.538 \\
0.643 & 0.651 & 0.848 & 0.97 & 0.744 & 0.845 & 0.581 & 0.483 & 0.499 & 0.609 & 0.729 & 0.627 & 0.702 & 0.791 & 0.146 & 0.964 & 0.612 & 0.425 & 0.727 & 0.748 & 0.565 \\ 
0.663 & 0.679 & 0.81 & 0.973 & 0.742 & 0.789 & 0.651 & 0.532 & 0.447 & 0.671 & 0.716 & 0.597 & 0.731 & 0.829 & 0.138 & 0.947 & 0.528 & 0.424 & 0.682 & 0.747 & 0.577 \\
0.69 & 0.696 & 0.877 & 0.981 & 0.7 & 0.787 & 0.583 & 0.437 & 0.42 & 0.598 & 0.698 & 0.633 & 0.69 & 0.791 & 0.159 & 0.948 & 0.522 & 0.441 & 0.736 & 0.72 & 0.503 \\ \hline
\end{tabular}
\end{table*}

\begin{table*}[!t]
\tiny \centering
\caption{Weakly Labelled SED audio event specific results for snr = 10} 
\label{table:class_sed_results_10}
\begin{tabular}{ c | c | c | c | c | c | c | c | c | c | c | c | c | c | c | c | c | c | c | c | c }
\hline
Model & \makecell{Guit\\ar} & \makecell{Appl\\ause} & \makecell{Ba\\rk} & \makecell{Bass\\drum} & \makecell{Bur\\ping} & \makecell{Bus} & \makecell{Cel\\lo} & \makecell{Chi\\me} & \makecell{Clar\\inet} & \makecell{Comp.\\keyb.} & \makecell{Cou\\gh} & \makecell{Cow\\bell} & \makecell{Double\\bass} & \makecell{Dra\\wer} & \makecell{Elec.\\piano} & \makecell{Fa\\rt} & \makecell{Finger\\snapp.} & \makecell{Fire\\work} & \makecell{Flu\\te} & \makecell{Glock\\ensp.}\\ \hline
GAP & 0.69 & 0.974 & 0.691 & 0.238 & 0.642 & 0.373 & 0.57 & 0.763 & 0.372 & 0.648 & 0.529 & 0.507 & 0.394 & 0.438 & 0.447 & 0.573 & 0.461 & 0.481 & 0.391 & 0.644 \\
GMP & 0.604 & 0.691 & 0.626 & 0.732 & 0.63 & 0.163 & 0.494 & 0.508 & 0.581 & 0.399 & 0.284 & 0.862 & 0.421 & 0.083 & 0.414 & 0.267 & 0.667 & 0.386 & 0.528 & 0.881 \\
GWRP & 0.777 & 0.969 & 0.868 & 0.454 & 0.873 & 0.49 & 0.685 & 0.809 & 0.597 & 0.668 & 0.766 & 0.842 & 0.512 & 0.553 & 0.527 & 0.665 & 0.567 & 0.643 & 0.552 & 0.921 \\
AtrousAP & 0.816 & 0.982 & 0.893 & 0.38 & 0.908 & 0.459 & 0.72 & 0.81 & 0.66 & 0.713 & 0.815 & 0.349 & 0.431 & 0.587 & 0.539 & 0.768 & 0.738 & 0.656 & 0.622 & 0.931 \\
2APAE & 0.921 & 0.953 & 0.861 & 0.904 & 0.942 & 0.672 & 0.775 & 0.674 & 0.583 & 0.728 & 0.747 & 0.894 & 0.78 & 0.614 & 0.652 & 0.652 & 0.954 & 0.73 & 0.69 & 0.85 \\
2APAE e-3 & 0.891 & 0.963 & 0.874 & 0.87 & 0.906 & 0.815 & 0.792 & 0.725 & 0.67 & 0.726 & 0.741 & 0.922 & 0.766 & 0.565 & 0.689 & 0.571 & 0.896 & 0.618 & 0.672 & 0.932 \\
2APAE e-2 & 0.856 & 0.937 & 0.816 & 0.905 & 0.884 & 0.715 & 0.762 & 0.565 & 0.659 & 0.684 & 0.627 & 0.918 & 0.805 & 0.516 & 0.673 & 0.647 & 0.911 & 0.716 & 0.696 & 0.929 \\ \hline
\end{tabular}
\end{table*}
\begin{table*}[!t]
\tiny \centering
\begin{tabular}{c | c | c | c | c | c | c | c | c | c | c | c | c | c | c | c | c | c | c | c | c}
\hline
\makecell{Gong} & \makecell{Gun\\shot} & \makecell{Harm\\onica} & \makecell{Hi-\\hat} & \makecell{Keys} & \makecell{Kno\\ck} & \makecell{Laugh\\ter} & \makecell{Me\\ow} & \makecell{Micro.\\oven} & \makecell{Oboe} & \makecell{Saxo\\phone} & \makecell{Scis\\sors} & \makecell{Shat\\ter} & \makecell{Snare\\drum} & \makecell{Squ\\eak} & \makecell{Tamb\\ourine} & \makecell{Tear\\ing} & \makecell{Tele\\phone} & \makecell{Trum\\pet} & \makecell{Violin\\fiddle} & \makecell{Writ\\ing} \\ \hline
0.513 & 0.537 & 0.854 & 0.802 & 0.486 & 0.579 & 0.569 & 0.437 & 0.261 & 0.403 & 0.554 & 0.538 & 0.645 & 0.617 & 0.219 & 0.598 & 0.422 & 0.396 & 0.728 & 0.553 & 0.499 \\
0.494 & 0.539 & 0.487 & 0.915 & 0.193 & 0.563 & 0.507 & 0.444 & 0.266 & 0.58 & 0.614 & 0.362 & 0.237 & 0.645 & 0.097 & 0.854 & 0.353 & 0.408 & 0.654 & 0.725 & 0.312 \\
0.684 & 0.705 & 0.884 & 0.913 & 0.679 & 0.714 & 0.71 & 0.684 & 0.514 & 0.541 & 0.751 & 0.597 & 0.77 & 0.763 & 0.246 & 0.873 & 0.506 & 0.569 & 0.819 & 0.712 & 0.534 \\
0.805 & 0.732 & 0.918 & 0.929 & 0.776 & 0.793 & 0.783 & 0.734 & 0.242 & 0.716 & 0.795 & 0.756 & 0.826 & 0.746 & 0.289 & 0.918 & 0.594 & 0.644 & 0.862 & 0.834 & 0.662 \\
0.656 & 0.749 & 0.869 & 0.975 & 0.773 & 0.849 & 0.675 & 0.481 & 0.534 & 0.659 & 0.765 & 0.604 & 0.723 & 0.873 & 0.175 & 0.943 & 0.593 & 0.506 & 0.803 & 0.727 & 0.596 \\
0.778 & 0.733 & 0.863 & 0.99 & 0.809 & 0.845 & 0.646 & 0.477 & 0.475 & 0.792 & 0.777 & 0.597 & 0.828 & 0.903 & 0.127 & 0.968 & 0.586 & 0.562 & 0.847 & 0.789 & 0.613 \\
0.659 & 0.72 & 0.851 & 0.983 & 0.774 & 0.83 & 0.661 & 0.445 & 0.527 & 0.698 & 0.747 & 0.552 & 0.796 & 0.947 & 0.127 & 0.941 & 0.628 & 0.522 & 0.813 & 0.775 & 0.571 \\ \hline
\end{tabular}
\end{table*}

\begin{table*}[!t]
\tiny \centering
\caption{Weakly Labelled SED audio event specific results for snr = 20} 
\label{table:class_sed_results_20}
\begin{tabular}{ c | c | c | c | c | c | c | c | c | c | c | c | c | c | c | c | c | c | c | c | c }
\hline
Model & \makecell{Guit\\ar} & \makecell{Appl\\ause} & \makecell{Ba\\rk} & \makecell{Bass\\drum} & \makecell{Bur\\ping} & \makecell{Bus} & \makecell{Cel\\lo} & \makecell{Chi\\me} & \makecell{Clar\\inet} & \makecell{Comp.\\keyb.} & \makecell{Cou\\gh} & \makecell{Cow\\bell} & \makecell{Double\\bass} & \makecell{Dra\\wer} & \makecell{Elec.\\piano} & \makecell{Fa\\rt} & \makecell{Finger\\snapp.} & \makecell{Fire\\work} & \makecell{Flu\\te} & \makecell{Glock\\ensp.}\\ \hline
GAP & 0.72 & 0.986 & 0.747 & 0.399 & 0.699 & 0.56 & 0.64 & 0.803 & 0.485 & 0.707 & 0.571 & 0.554 & 0.501 & 0.532 & 0.597 & 0.652 & 0.481 & 0.593 & 0.498 & 0.766\\
GMP &  0.507 & 0.843 & 0.654 & 0.838 & 0.631 & 0.336 & 0.565 & 0.489 & 0.657 & 0.344 & 0.44 & 0.889 & 0.42 & 0.137 & 0.579 & 0.328 & 0.653 & 0.226 & 0.54 & 0.931\\
GWRP & 0.83 & 0.986 & 0.922 & 0.529 & 0.869 & 0.649 & 0.727 & 0.813 & 0.657 & 0.728 & 0.742 & 0.875 & 0.696 & 0.626 & 0.627 & 0.7 & 0.636 & 0.722 & 0.697 & 0.934\\
AtrousAP & 0.877 & 0.991 & 0.922 & 0.562 & 0.924 & 0.622 & 0.773 & 0.819 & 0.746 & 0.77 & 0.89 & 0.716 & 0.573 & 0.708 & 0.703 & 0.806 & 0.746 & 0.755 & 0.745 & 0.957\\
2APAE & 0.903 & 0.969 & 0.911 & 0.936 & 0.959 & 0.761 & 0.787 & 0.642 & 0.666 & 0.736 & 0.605 & 0.936 & 0.825 & 0.592 & 0.665 & 0.589 & 0.956 & 0.681 & 0.834 & 0.913\\
2APAE e-3 & 0.908 & 0.955 & 0.867 & 0.9 & 0.946 & 0.755 & 0.845 & 0.648 & 0.738 & 0.716 & 0.707 & 0.936 & 0.81 & 0.611 & 0.708 & 0.682 & 0.909 & 0.712 & 0.849 & 0.961\\
2APAE e-2 & 0.89 & 0.97 & 0.908 & 0.922 & 0.93 & 0.737 & 0.874 & 0.598 & 0.647 & 0.643 & 0.668 & 0.946 & 0.826 & 0.656 & 0.667 & 0.583 & 0.931 & 0.707 & 0.74 & 0.84\\ \hline 
\end{tabular}
\end{table*}
\begin{table*}[!t]
\tiny \centering
\begin{tabular}{ c | c | c | c | c | c | c | c | c | c | c | c | c | c | c | c | c | c | c | c | c }
\hline
\makecell{Gong}  & \makecell{Gun\\shot} & \makecell{Harm\\onica} & \makecell{Hi-\\hat} & \makecell{Keys} & \makecell{Kno\\ck} & \makecell{Laugh\\ter} & \makecell{Me\\ow} & \makecell{Micro.\\oven} & \makecell{Oboe} & \makecell{Saxo\\phone} & \makecell{Scis\\sors} & \makecell{Shat\\ter} & \makecell{Snare\\drum} & \makecell{Squ\\eak} & \makecell{Tamb\\ourine} & \makecell{Tear\\ing} & \makecell{Tele\\phone} & \makecell{Trum\\pet} & \makecell{Violin\\fiddle} & \makecell{Writ\\ing} \\ \hline
0.664 & 0.552 & 0.895 & 0.829 & 0.535 & 0.688 & 0.642 & 0.501 & 0.308 & 0.548 & 0.622 & 0.554 & 0.717 & 0.696 & 0.236 & 0.67 & 0.432 & 0.445 & 0.825 & 0.626 & 0.531\\
0.529 & 0.523 & 0.517 & 0.804 & 0.335 & 0.599 & 0.441 & 0.302 & 0.121 & 0.668 & 0.653 & 0.371 & 0.289 & 0.509 & 0.076 & 0.873 & 0.4 & 0.404 & 0.72 & 0.726 & 0.277\\
0.811 & 0.713 & 0.917 & 0.92 & 0.733 & 0.827 & 0.791 & 0.674 & 0.498 & 0.728 & 0.801 & 0.624 & 0.812 & 0.79 & 0.277 & 0.915 & 0.571 & 0.647 & 0.869 & 0.839 & 0.65\\
0.86 & 0.743 & 0.945 & 0.954 & 0.782 & 0.832 & 0.776 & 0.724 & 0.123 & 0.84 & 0.862 & 0.74 & 0.859 & 0.762 & 0.239 & 0.955 & 0.613 & 0.646 & 0.909 & 0.888 & 0.686\\
0.751 & 0.676 & 0.891 & 0.966 & 0.79 & 0.838 & 0.637 & 0.498 & 0.552 & 0.79 & 0.833 & 0.633 & 0.846 & 0.932 & 0.323 & 0.965 & 0.609 & 0.503 & 0.903 & 0.855 & 0.566\\
0.725 & 0.666 & 0.87 & 0.989 & 0.76 & 0.808 & 0.705 & 0.455 & 0.616 & 0.83 & 0.833 & 0.613 & 0.809 & 0.943 & 0.177 & 0.966 & 0.56 & 0.495 & 0.885 & 0.819 & 0.578\\
0.719 & 0.707 & 0.849 & 0.971 & 0.82 & 0.869 & 0.674 & 0.436 & 0.532 & 0.735 & 0.817 & 0.546 & 0.781 & 0.904 & 0.182 & 0.967 & 0.566 & 0.465 & 0.83 & 0.834 & 0.6\\ \hline
\end{tabular}
\end{table*}

\subsection{Ablation Study \label{section:results_ablationstudy}}
To determine the contribution of 2-step Attention Pooling and reconstruction based auxiliary task on the SED performance, ablation study is performed. For ablation study, we change the value of $\alpha$ in total loss, described in Section \ref{section:proposedsolution_formulation}:
\begin{align}
    \mathcal{L} = \mathcal{L}_1(P, y \vert w, w_4) + \alpha \mathcal{L}_1(\{\bar{x_i}\}_{i = 1}^T, \{\hat{x_i}\}_{i = 1}^T \vert w, w_2)
\end{align}
The value of $\alpha$ determines the contribution of auxiliary task to the weakly labelled SED. By varying the value of alpha, we can determine the performance improvement by 2-step attention ($\alpha$ = 0.0) and contribution of auxiliary task ($\alpha >$ 0). The alpha values are kept low as the $\mathcal{L}_2$ loss term (Mean Squared Error) is magnitude greater than $\mathcal{L}_1$ loss. The lower $\alpha$ value helps in adjusting for this large scale difference between the two losses.

When $\alpha = 0.0$, the network has no contribution from the auxiliary task and can be used to evaluate the performance of 2-step Attention Pooling. In terms of micro-Precision, the 2-step Attention Pooling outperforms existing benchmark of Atrous AP (Atrous Attention Pooling \cite{atrous}) from Table \ref{table:ablation_study_results} on SNR 20, 10 and 0 dB by 5.2\%, 10.2\% and 21.4\% respectively. By adding the auxiliary task contribution with a relative weightage of $\alpha = 0.001$, compared to proposed 2-step attention, an improvement of 0.7\%, 2.3\% and 0.7\% is observed. The numbers indicate that 2-step attention has a heavy contribution on the improvement of performance, with extra performance gains from auxiliary task. When $\alpha$ is increased to 0.01, the performance compared to $\alpha = 0.001$ is decreased. This indicates that the auxiliary task's loss contribution starts to overpower the primary SED task's loss contribution rather than improving generalisation. The shared encoder then learns features more relevant to auxiliary task rather than primary SED task. 

\begin{figure*}[!t]
    \centering
    \hspace*{-0.5cm}\includegraphics[width=6.5 in, height = 5in]{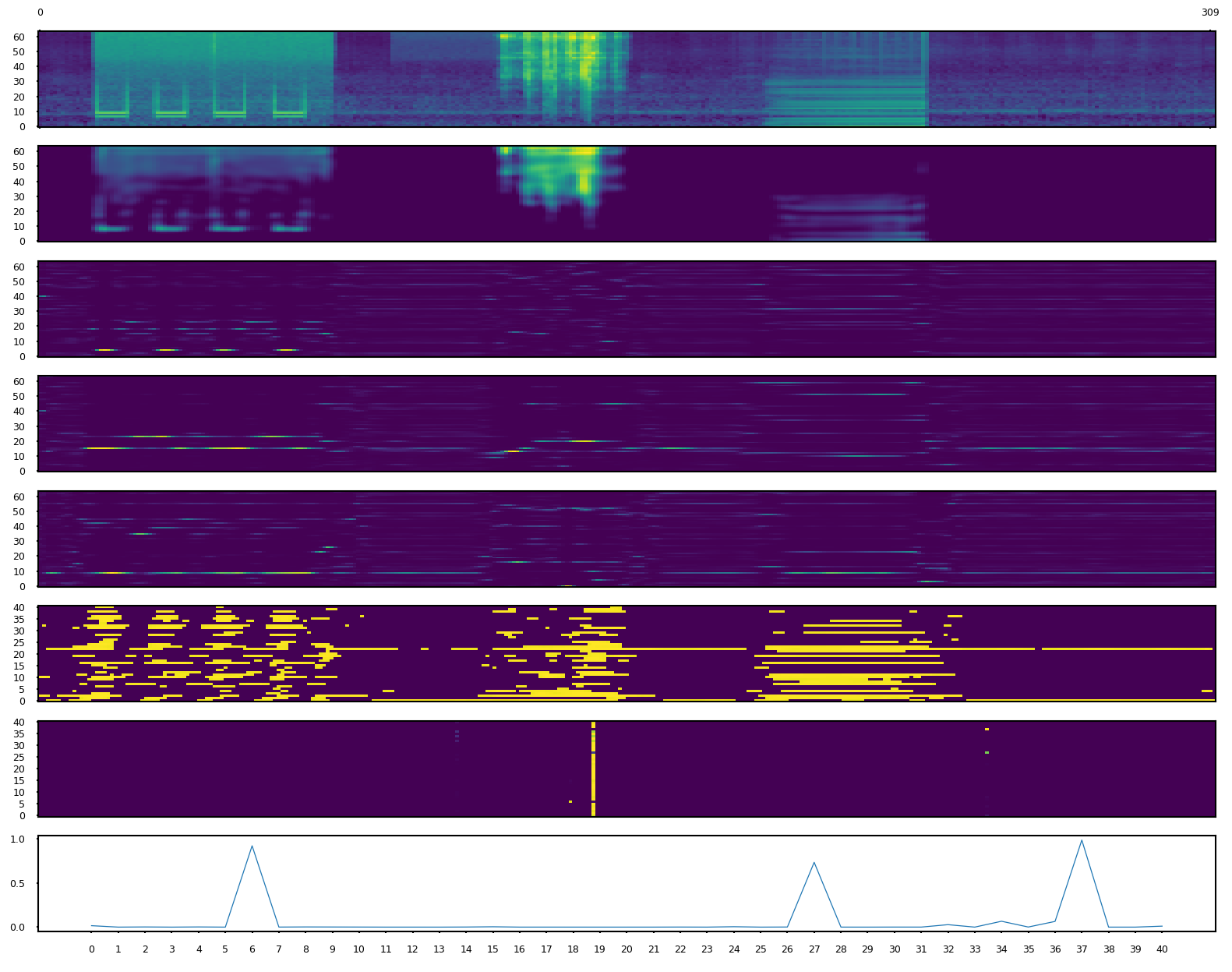}
    \caption{Visualisation of 2 step Attention Pooling and Auxiliary task's decoder outputs. The figure has total of 8 subplots each visualising a step of the 2APAE process. Subplot 1 depicts the scaled log mel spectrogram input to the network. Subplot 2 is the reconstructed output by the auxiliary task decoder. Subplots 3,4 and 5 are attention weights for three most probable classes in the prediction. Subplot 6 is the output of $1^{st}$ step Attention pooling. Subplot 7 and 8 is the attention weight and output of $2^{nd}$ step Attention Pooling respectively. The y-axis in subplot 1 - 4 is Mel-bins and sound event in subplot 5-6. The x-axis is time in subplot 1-6 and Sound Event in subplot 7}
    \label{fig:visualisation}
\end{figure*}

\subsection{Audio-event specific SED results}
Table \ref{table:class_sed_results_0}, \ref{table:class_sed_results_10}, \ref{table:class_sed_results_20} shows precision values for audio event specific SED for each of the 41 audio events for SNR = 0,10 and 20 dB respectively. The name of the audio events are abbreviated to fit the table. In the table, 2APAE refers to the base model with no contribution from auxiliary task, while 2APAE 1e-3 and 2APAE 1e-3 refers to 2APAE model with auxiliary task contribution of $\alpha = 0.001$ and $\alpha = 0.01$ respectively. For almost all audio events, variants of 2APAE have the best precision scores against GMP, GAP, GWRP, Atrous for all snr = 0, 10, 20. Particularly, for audio events like 'Bass drum', 'bus', 'double bass', 'cowbell' 2APAE outperforms other models by a large margin. However, 2APAE struggles achieving best scores for audio events like 'gong','chime' and 'meow' where the Atrous model performs better. This indicates using atrous or dilated convolutions can be helpful in particular situations like detecting sound events whose energy is spread wide in the temporal domain. This can be incorporated in 2APAE by replacing the linear convolutions in shared encoder with atrous or dilated convolutions. This direction is not explored in this paper and is left for future work.



\subsection{Interpretable visualisation \label{section:results_vis}}
Apart from improved performance, using 2-step Attention Pooling, provides a way to visualise the contribution of each Mel bin and each time frame in the T-F representation to each audio event and final predictions. Before visualising attention weights, generally a class-dependent threshold is applied \cite{gwrp} \cite{mit} to remove False positive and fake activations. But this post-processed varies from network to network and doesn't paint an accurate picture of the contribution of each component in attention as it becomes highly dependent on threshold chosen and adds author's bias for choosing threshold. Hence in this paper we plot the raw attention weights without applying any thresholds. 

We pick a random example with SNR 20 db and show the end to end visualisation of the 2 Step Attention Pooling mechanism. The first subplot shows the log mel spectrogram that is fed to the segmentation network. The first subplot has three distinct audio events happening in it, namely: telephone ringing, cello playing and cat meowing. The second subplot shows the output of reconstruction based auxiliary tasks. We will get back to the description of the second subplot later. The third, fourth, fifth subplot shows the $1^{st}$ Step Attention Pooling weights for top 3 predicted events by the network: telephone ringing, cello playing and cat meowing respectively with time on x-axis and mel bins on y-axis. The lighter color indicates higher weights in time and frequency domain. These subplots tell us what frequency bands are getting activated at which instance of time for the particular class. For example, in subplot 3, the lower frequency bands are getting activated for initial time steps where the telephone ringing is occurring in the audio clip. The fifth subplot shows the output of $1^{st}$ step Attention Pooling where y-axis is audio events and x-axis is time with squashed frequency axis. This shows that after taking into account the contribution from every frequency bin, there are three main candidate time chunks where the audio events are located. Comparing this to subplot one of input, they align with the three audio events happening in the audio clip, with different audio events getting activated at different time chunks. The sixth subplot visualises the $2^{nd}$ Step Attention weights where y-axis is audio events and x-axis is time. They give high weight to audio events telephone ringing, cello playing and cat meowing at different time steps. However, the audio events are not perfectly time-aligned. This is due to the use of 2D-Average pooling per two convolution layers in the shared encoder. The addition of 2D average pooling improves weakly labelled SED performance, but results in losing time-level precision. The seventh subplot, shows the output of $2^{nd}$ step Attention Pooling where y-axis is the audio event prediction probability and the x-axis is the audio event. This is the final output of the classification network and entire network for SED. 

Coming back to subplot two, it is the output of the decoder used for reconstruction auxiliary task. In our case, the goal of the auxiliary task and the decoder is to reconstruct the log mel spectrogram of input from the shared encoder's T-F representation. From the subplot, we can see that the decoder is not only able to reconstruct the audio events clearly but it is also denoising the log mel spectrogram in 20 dB SNR. For the decoder to perform well and denoise the output, the internal T-F representation has to be denoised as well. This helps us in inferring two things: first, In Multi-Task Learning framework, adding the reconstruction auxiliary task to primary weakly labelled SED leads to denoising of the internal T-F and the encoder representation of log mel spectrogram. This is backed by ablation study's result in Section \ref{section:results_ablationstudy} which shows adding reconstruction auxiliary task improves performance. That improved performance can be attributed to this denoising of internal T-F representations. Second, the denoised log mel representation, will help in improving the performance of audio source separation systems. This setup either pre-trained or jointly trained with source separation, can be used to produce cleaned audio log mel representation for better source separation.

\section{Future work and directions \label{section:futurework}}
The paper shows, Weakly labelled SED can be formulated as Multi-Task Learning where Multiple Instance Learning SED is the primary task coupled with an constructive auxiliary task. This results in interesting directions for extending and building upon this work. To enumerate a few: first, exploring different auxiliary task to help Weakly labelled SED detection. The paper shows one such task and there can be single or multiple auxiliary task which might help creating better shared segmentation mapping for audio events. Second, jointly training sound source separation along with weakly Labelled SED in MTL framework. This would provide an end-to-end weakly labelled sound event detection and source separation system. Third, developing better loss function and dynamic $\alpha$ computation to optimally use multiple audio task in MTL framework for improved performance.

\section{Conclusion \label{section:conclusion}}
The paper proposes a Multi-Task learning formulation for learning from Weakly Labelled Audio data which incorporates the MIL formulation of SED as primary task. In the MTL framework, we use input T-F reconstruction as auxiliary task which helps in denoising the intermediate T-F representation and encourages the shared segmentation network to retain source specific information. To make the pooling mechanism more interpretable in T-F domains, we introduce a 2-step Attention Pooling mechanism. To show the utility of the proposed framework, we remix the DCASE 2019 task 1 acoustic scene data with DCASE 2018 Task 2 sounds event data under 0, 10 and 20 db SNR. The proposed network outperforms existing benchmark model over all SNRs, specifically 22.3 \%, 12.8 \%, 5.9 \% improvement over benchmark model on 0, 10 and 20 dB SNR respectively. The result from the ablation study indicates: the 2-step Attention Pooling is the major factor improving performance of SED followed by about 1\% improvement by addition of reconstruction based auxiliary task. The work lays foundation for future work of end-to-end sound source separation and sound event detection for Weakly Labelled data.


%





\ifCLASSOPTIONcaptionsoff
  \newpage
\fi



%
\bibliographystyle{IEEEtran}
\bibliography{main}

%








\end{document}